\documentclass[epj]{webofc}
\usepackage[varg]{txfonts}   
\usepackage{color}
\usepackage{amsmath,amssymb,cancel}
\usepackage{stackrel,enumerate}
\usepackage{graphicx,hyperref}
\usepackage{slashed, mathrsfs}
\usepackage{youngtab}
\usepackage{relsize}

\newcommand{\e}{\textrm{e}}
\newcommand{\state}[1]{\left|#1\right>}

\newcommand{\bstate}[1]{\left<\bar{#1}\right|}

\makeatletter
\DeclareRobustCommand{\rvdots}{%
  \vbox{
    \baselineskip4\p@\lineskiplimit\z@
    \vskip0.1em
    \hbox{.}\hbox{.}
  }}
  
\makeatother
\def\Tiny{\fontsize{3.5pt}{3.5pt} \selectfont}
\def\Small{\fontsize{6pt}{6pt} \selectfont}

\newcommand{\tr}{\textrm{tr}}

\wocname{EPJ Web of Conferences}
\woctitle{ICNFP 2017}
\begin{document}
\selectlanguage{english}
\title{Worldline colour fields and non-Abelian quantum field theory}
%
%

\author{James P. Edwards\inst{1}\fnsep\thanks{\email{jedwards@ifm.umich.mx}} \and Olindo Corradini \inst{2,3}\fnsep\thanks{\email{olindo.corradini@unimore.it}}}

\institute{Instituto de F\'isica y Matem\'aticas,
Universidad Michoacana de San Nicol\'as de Hidalgo,
Edificio C-3, Apdo. \!Postal 2-82,
C.P. 58040, Morelia, Michoac\'an, Mexico \and Dipartimento di Scienze Fisiche, Informatiche e Matematiche, Universit\`a degli Studi di Modena e Reggio Emilia, Via Campi 213/A, I-41125 Modena, Italy \and INFN, Sezione di Bologna, Via Irnerio 46, I-40126 Bologna, Italy}

\abstract{In the worldline approach to non-Abelian field theory the colour degrees of freedom of the coupling to the gauge potential can be incorporated using worldline ``colour'' fields. The colour fields generate Wilson loop interactions whilst Chern-Simons terms project onto an irreducible representation of the gauge group. We analyse this augmented worldline theory in phase space focusing on its supersymmetry and constraint algebra, arriving at a locally supersymmetric theory in superspace. We demonstrate canonical quantisation and the path integral on $S^{1}$ for simple representations of $S\!U(N)$.}
\maketitle

\section{Introduction}
Non-Abelian symmetry groups are a crucial aspect of theoretical physics, with focus on groups such as $S\!U(2)$, $S\!U(3)$ of the standard model and $S\!U(5)$ or $S\!O(10)$ that appear in attempts at unification \cite{Kibble, tHooft}. New tools for the study of non-Abelian field theory would be welcome if they were to shed light on the analytic calculation of scattering amplitudes or the non-perturbative phenomenon of confinement. One such tool that is gaining in popularity is the \textit{worldline formalism} of quantum field theory \cite{Strass1, ChrisRev, 3App}. Its origins can be traced to Feynman \cite{FeynEM}, with a revival of interest following Bern and Kosower's Master Formulae for field theory amplitudes derived from string theory \cite{BK1, BK2}. It is recently finding application in a wide range of physical problems, including multi-loop effective actions and scattering amplitudes \cite{Strass1, Super1}, constant field QED and the Schwinger effect \cite{Ffield1, Ffield2}, gravitational interactions \cite{Grav1, Grav2}, higher spin fields \cite{HSWL1, HSWL2}, quantum fields in non-commutative space-time \cite{NCWL1, NCWL2} and the form factor decomposition of the four gluon vertex reported by Schubert elsewhere in these proceedings. 

Worldline techniques provide a first quantised approach to studying quantum field theory, where the main object to calculate becomes a path integral over point particle trajectories. This is supplemented by functional integrals over additional variables on the particle worldlines, representing further degrees of freedom such as spin, Faddeev-Popov ghosts etc. This approach has been adapted to the study of non-Abelian field theory, using worldline \textit{colour fields} to represent gauge group degrees of freedom \cite{Col1, Col2, MeUnif, Paul, ColTree}, later generalised to matter transforming in arbitrary representations in \cite{JO1, JO2}. Here we recap this approach and extend previous discussions on the worldline supersymmetry of the colour-field augmented worldline action. We first examine the phase-space worldline action for a free particle and then couple it to a non-Abelian background, reformulating the action in super-space where its supersymmetry becomes manifest. This will uncover a $U(F)$ worldline symmetry which we will gauge, showing how the resulting first class constraints project onto a subspace of the particle Hilbert space meaning it transforms in an arbitrary irreducible representation of the gauge group. Thus any matter multiplet can be described with a single worldline theory without the need for a manual path ordering prescription, that is instead generated automatically by the path integral over the colour fields. This procedure leads to compact expressions in perturbation theory that combine many Feynman diagrams through preservation of permutation symmetry of external gluon legs.

\section{Locally supersymmetric spinning particle coupled to a non-Abelian field}
We start from a free, locally symmetric, $\mathcal{N}=1$ spinning particle. The phase space action reads \cite{BdVH}
\begin{align}
S[x,p,\psi] =\int_0^1d\tau \Big[ p\cdot \dot x +\frac{i}{2}\psi\cdot \dot \psi-e H -i\chi Q\Big]  
\label{SFree}
\end{align}
where
\begin{align}
\qquad H =\frac12 p^2; \qquad Q =\psi^\mu p_\mu
\end{align}
are the Hamiltonian and the SUSY charge respectively. These generate a supersymmetry and translation invariance that are gauged by $e(\tau)$, the einbein, and $\chi(\tau)$ the (Grassmann) gravitino. This action appears in the worldline approach to the quantisation of a free Dirac field at one-loop order through a point particle path integral representation of the partition function \cite{Strass1}\footnote{The worldline approach to spinor QED is based on the second-order formalism \cite{SO1, SO2} as discussed in \cite{Morgan}.}
\begin{equation}
\ln\mathscr{Z} = \ln\textrm{Det}^{-\frac{1}{2}}\left[\left(\gamma \cdot \partial\right)^{2} + m^2\right] = -\frac{1}{2}\int_{0}^{\infty}\frac{dT}{T} \e^{-\frac{i}{2}m^{2}T}\oint_{x(0) = x(T)} \hspace{-2em} \mathscr{D}x\,\, \oint_{\psi(0) + \psi(T) = 0} \hspace{-2em}\mathscr{D}\psi\,\,\int \mathscr{D}p \, \e^{iS_{\textrm{gf}}[x,p,\psi]}.
	\label{WL}
\end{equation}
The bosonic coordinates $x^{\mu}$ represent an embedding of the worldline trajectory in phase space (with momenta $p^{\mu}$) whilst the $\psi^{\mu}$ generate the spin information of the field; they are respectively periodic and anti-periodic on $S^{1}$. The action (\ref{SFree}) has been gauge fixed to $\chi = 0$ and $e = T$ which introduces a Faddeev-Popov determinant $\frac{1}{T}$ as a measure on the integral over the Schwinger proper time \cite{Schwinger} in which the particle traverses the closed loops (the proper time integral produces the logarithm on the left hand side whilst the boundary conditions implement the functional trace following from $\ln \textrm{Det} = \textrm{Tr} \ln$. The mass term in the exponent can be generated by adding $\frac{m^{2}}{2}$ to the canonical momentum). 

The (Dirac)-Poisson brackets follow from the symplectic terms of the action and are
\begin{align}
\{ x^\mu,p_\nu\} =\delta^\mu_\nu; \qquad \{\psi^\mu,\psi^\nu\} =-i\eta^{\mu\nu}
\end{align}
which implies a closed supersymmetry algebra
 \begin{align}
 \{Q,Q\} =-2iH; \qquad \{Q,H\} =0; \qquad \{H, H\} = 0.
 \label{eq:QQ}
 \end{align}
The infinitesimal transformations of the fields follow from $\delta \phi = \{\phi, \xi(\tau)H + i \eta(\tau)Q\}$ and read
\begin{align}
\begin{split}
\delta_\xi x^\mu  = \xi p^\mu; \qquad \delta_\eta x^\mu &=i\eta \psi^\mu; \qquad\delta_\xi p_\mu =0;\qquad \delta_\eta p_\mu=0\\
 \delta_\xi \psi^\mu & =0;\qquad \quad \hspace{0.25em} \delta_\eta \psi^\mu=-\eta p^\mu\\
\delta_\xi \chi = 0\;\qquad \hspace{1.5em}\delta_\eta \chi &= \dot\eta; \qquad \hspace{1.9em} \delta_\xi e = \dot \xi; \qquad \hspace{0.5em}\delta_\eta e = 2i\chi \eta 
\end{split}
\end{align}
\subsection{Non-Abelian field theory}
Let us now try to couple the particle to a non-Abelian gauge field. We start with a naive covariantisation of the momentum (we absorb the coupling strength into the gauge potential $A_{\mu}$)
\begin{align*}
p_\mu \ \longrightarrow \ \pi_\mu =p_\mu- A_\mu\,,\quad Q:= \psi^\mu \pi_\mu
\end{align*}
with $A_\mu:= A_\mu^a T^a \in su(N)$ with $T^a$ written in some fixed representation, that for definiteness we take as the fundamental representation. We would also require a path ordering prescription in (\ref{WL}) to take into account that the gauge group generators do not in general commute. Thus, using the above Poisson brackets, we get the algebra~\eqref{eq:QQ} where the Hamiltonian is now given by 
\begin{align}
H= \frac12 \pi^2 +\frac{i}{2} \psi^\mu F_{\mu\nu}\psi^\nu
\end{align} 
with $F=d\wedge A$ being the \textit{Abelian part} of the field strength $G=d\wedge A -iA\wedge A$ associated to $A$. 

One possibility for generating the full field strength from the Poisson bracket is application of the super-space formalism and the definition of a super path ordering (see \cite{ChrisRev} or the appendix of \cite{JO2}). Here, however, we follow \cite{Col1, JO1} and add $N$ dynamical auxiliary {\it colour} fields, $\bar c^\alpha, c_\alpha $ that transform in the (anti)fundamental representation of the gauge group -- they can either be commuting or anti-commuting and here we consider the anti-commuting version. Then we define 
\begin{align}
A_\mu \ \longrightarrow \ {\cal A}_\mu := \bar c^\alpha A_\mu^a (T^a)_\alpha{}^{\alpha'} c_{\alpha'}
\end{align}  
and construct the covariant momentum, Hamiltonian and super-charge with ${\cal A}_\mu$. Furthermore, choosing the action for the auxiliary variables as $\int d\tau \,i\bar c^\alpha \dot c_\alpha$, the associated Poisson brackets read
\begin{align}
\{\bar c^\alpha\, c_{\alpha'}\} =-i\delta^\alpha_{\alpha'}
\end{align}
that mean that the colour fields provide a classical representation of the Lie algebra: define $R^{a} = \bar{c}^{\alpha}(T^{a})_{\alpha}{}^{\alpha'}c_{\alpha'}$ and one finds that that $\{R^{a}, R^{b}\} = f^{abc}R^{c}$. This in turn implies the first class algebra ~\eqref{eq:QQ} where now the generators of reparameterisations and supersymmetry are given by
\begin{equation}
	\widetilde{H} = \frac{1}{2}\widetilde{\pi}^{2} + \frac{i}{2}\psi^{\mu}\mathcal{G}_{\mu\nu}\psi^{\nu}; \qquad \widetilde{Q} = \tilde{\pi} \cdot \psi; \qquad \widetilde{\pi}^{\mu} = p^{\mu} - \bar{c}^{\alpha \prime} (A^{\mu})_{\alpha \prime}{}^{\alpha} c_{\alpha},
\end{equation}
which provide us with a representation of the SUSY algebra
\begin{equation}
	\left\{\widetilde{Q}, \widetilde{Q}\right\} = -2i\widetilde{H}.
\end{equation}
with a Hamiltonian where $F_{\mu\nu}$ is completed to the full field strength tensor through
\begin{align}
&{\cal G}_{\mu\nu} = \bar c^\alpha ( G_{\mu\nu})_\alpha{}^{\alpha'} c_{\alpha'}; \qquad ( G_{\mu\nu})_\alpha{}^{\alpha'}= i\big[\partial_\mu-i{ A}_\mu,  \partial_\nu-i{ A}_\nu\big] = \big( F_{\mu\nu}-i\big[A_\mu,A_\nu \big]\big)_\alpha{}^{\alpha'} .
\end{align}
Moreover, the symplectic structure of the colour fields is sufficient for them to generate the required path ordering prescription that in previous worldline calculations has been put in by hand \cite{ChrisRev, PathO, Strass1}, since their worldline Green functions is essentially a step function. With such improvement on the covariant momentum the phase space action becomes
\begin{equation}
	S[x,p,\psi, \bar{c}, c] =\int_0^1d\tau \Big[ p\cdot \dot{x} +\frac{i}{2}\psi\cdot \dot{\psi} + i\bar{c}^{\alpha}\dot{c}_{\alpha} -e \widetilde{H} -i\chi \widetilde{Q} \Big]  
	\label{phasec}
\end{equation}
The transformations of the fields is determined by Poisson brackets with the new generators, which in particular means that now the auxiliary colour fields transform under the supersymmetry as
\begin{equation}
 \delta_\eta c_\alpha = \big\{ c_\alpha, i\eta \widetilde{Q}\big\} =-\eta \psi^\mu (A_\mu)_\alpha{}^{\alpha'} c_{\alpha'}; \qquad \delta_\eta \bar c^\alpha = \big\{ \bar c^\alpha, i\eta \widetilde{Q}\big\} = \eta \psi^\mu \bar c^{\alpha'} (A_\mu)_{\alpha'}{}^{\alpha}
 \label{delc}
\end{equation}
(note also that when the particle is coupled to the gauge field the momenta also change under supersymmetry as $\delta_{\eta} p_{\mu} = \{p_{\mu}, i\eta \widetilde{Q}\} = i\eta \psi^{\nu}\partial_{\mu} \bar{c}^{\alpha \prime}  (A_{\nu})_{\alpha \prime}{}^{\alpha} c_{\alpha}$).

\subsection{Canonical quantisation}
\label{canon}
As is well known \cite{BdVH, Deser} the equations of motion for $e$ and $\chi$ are imposed as constraints on physical states of the Hilbert space, providing respectively the mass shell condition and Dirac equation that decouple negative norm states. Furthermore the colour fields generate a Hilbert space populated by acting on the vacuum with creation operators $\bar{c}^{\alpha} \rightarrow \hat{c}^{\dagger \alpha}$. In a coherent state basis, we define 
\begin{equation}
	\bstate{u} = \langle 0|e^{\bar u^{r} \hat{c}_{r}}; \qquad \bstate{u}\hat{c}^{\dagger r} = \bar{u}^{r}\bstate{u}; \qquad \bstate{u}\hat{c}_{r} = \partial_{\bar{u}^{r}}\bstate{u},
\end{equation}
and find that the wavefunction is described by a reducible sum of components transforming in fully anti-symmetric representations of the gauge group (the sum is finite for Grassmann colour fields -- when using commuting colour fields one finds an infinite sum over fully \textit{symmetric} representations)
\begin{equation}
	\Psi(x, \bar{u}) = \psi(x) + \psi_{r_{1}}(x)\bar{u}^{r_{1}} + \psi_{r_{1} r_{2}}(x) \bar{u}^{r_{1}}\bar{u}^{r_{2}} + \ldots + \psi_{r_{1} r_{2} \ldots  r_{N}}(x)\bar{u}^{r_{1}}\bar{u}^{r_{2}}\cdots\bar{u}^{r_{N}}; \quad \Yvcentermath1
	\psi_{r_{1} r_{2} ... r_{p}} \sim \underbrace{\underset{ \,\Small\yng(1,1)\, }{\overset{ \,\Small\yng(1,1)\, } {\rvdots} }}_{p} \,, 
\Yvcentermath0	
	\label{Psi}
\end{equation}
where we have indicated the Young Tableau describing the representation in which each component transforms. So to describe a multiplet transforming in an irreducible representation it has become necessary to project onto the appropriate subspace. Moreover, as it stands this worldline theory only gives  access to fully anti-symmetric representations (sufficient for standard model multiplets and the $\mathbf{1}$, $\mathbf{\bar{5}}$ and $\mathbf{10}$ of $S\!U(5)$ unified theory as discussed in \cite{Paul, MeUnif}) and it is natural to extend this description to components with arbitrary symmetry. To this end, in \cite{JO1, JO2} the worldline action was extended to incorporate $F$-families of colour fields $\bar{c}^{\alpha}_{f}$, $c_{f \alpha}$ for $f \in \{1, \ldots F\}$. Then the worldline action becomes
\begin{equation}
	S\left[x, p, \psi, e, \chi, \bar{c}, c\right] = \int_{0}^{1} d\tau \, \bigg[ p \cdot \dot{x} + \frac{i}{2}\psi \cdot \dot{\psi} +i \bar{c}_{f}^r \dot{c}_{fr} - e\widetilde{H} - i \chi \widetilde{Q} \bigg],
	\label{Scf}
\end{equation}
where the symplectic term for the colour fields and the supersymmetry generators now involve sums over all $F$ families. In this case, the wavefunction now transforms as a sum over all possible $F$-fold tensor products of fully anti-symmetric representations of $S\!U(N)$:
\begin{equation}
\Yvcentermath1
	\Psi ~ \sim \sum_{\{p_{1}, p_{2}, \ldots p_{\!F}\}}  \underbrace{\underset{ \,\Small \yng(1,1)\, }{\overset{ \,\Small\yng(1,1)\, } {\rvdots} }}_{p_{F}} \otimes  \ldots \otimes \underbrace{\underset{ \,\Small\yng(1,1)\, }{\overset{ \,\Small\yng(1,1)\, } {\rvdots} }}_{p_{2}}   \otimes \underbrace{\underset{ \,\Small\yng(1,1)\, }{\overset{ \,\Small\yng(1,1)\, } {\rvdots} }}_{p_{1}}  . 
	\Yvcentermath0
\end{equation}
We now explain how to select wavefunction components transforming in an arbitrary irreducible representation from this sum.

Note that the action (\ref{Scf}) has an additional global $U(F)$ symmetry rotating between the families of colour fields. If $\Lambda_{fg}$ is a constant element of $U(F)$ then the action is invariant under transformations 
\begin{equation}
	c_{fr} \rightarrow \Lambda_{fg} c_{gr}; \qquad \bar{c}_{f}^r \rightarrow \bar{c}_{g}^r \Lambda^\dagger_{gf}
\end{equation}
with corresponding generators $L_{fg} := \bar{c}^{\alpha}_{f}c_{g\alpha}$ that produce field variations through Poisson brackets. The generators satisfy the $U(N)$ algebra
\begin{equation}
	\{L_{fg}, L_{f\prime g\prime}\} = i\delta_{f g\prime}L_{f \prime g} - i\delta_{f \prime g}L_{f g \prime}\, .
\end{equation}
The identification of these charges allows us to gauge this symmetry for an action that is invariant under transformations whose parameters depend on worldline time. It is necessary to make a \textit{partial} gauging of the closed sub-algebra of generators $L_{fg}$ with $f \leqslant g$ leading to a new phase space action
\begin{equation}
	S[x,p,\psi, \bar{c}, c, \bar{z}, z, a] =\int_0^1d\tau \Big[ p\cdot \dot{x} +\frac{i}{2}\psi\cdot \dot{\psi} + \sum_{f = 1}^{F} i\bar{c}^{\alpha}_{f}\dot{c}_{f \alpha} -e \widetilde{H} -i\chi \widetilde{Q} - \sum_{f\leqslant g}a_{f g}\left(L_{fg} - s_{f}\delta_{fg}\right)\Big]  
	\label{phasecza}
\end{equation}
where the $a_{fg}(\tau)$ are the $U(F)$ worldline gauge fields that compensate for the variation of the colour fields under local transformations. Under the $U(F)$ symmetry these variables transform as
\begin{equation}
	a_{fg} \rightarrow \Lambda^{\dagger}_{fh} \dot{\Lambda}_{hg} + i\Lambda^{\dagger}_{fh} a_{hk} \Lambda_{kg}.
	\label{da}
\end{equation} 
Now the equations of motion for the worldline gauge fields will impose constraints on the physical states of the system. This will be sufficient to project onto a single irreducible representation of the gauge group, facilitated by the introduction of the non-minimal Chern-Simons terms $\sum_{f}s_{f}a_{ff}$ which are invariant under the partially gauged $U(F)$ symmetry (\ref{da}). These tunable parameters $s_{f} := n_{f} - \frac{N}{2}$ will allow us to fix the occupation number of family $f$ of the colour fields to be $n_{f}$.

In a coherent state basis the $U(F)$ generators become $\hat L_{fg} = \bar{u}^{\alpha}_{f} \partial_{\bar{u}^{\alpha}_{g}}$. The equations of motion for the worldline gauge fields $a_{fg}$ impose the following constraints on the state space:
\begin{align}
\left(\hat{L}_{ff} + \frac{N}{2}\right)\state{\Psi} &= n_{f}\state{\Psi} \longrightarrow \left(\bar{u}^{\alpha}_{f}\frac{\partial}{\partial \bar{u}_{f}^{\alpha}} - n_{f}\right)\Psi(x, \bar{u}) = 0  \label{ConDiag}\\
\hat{L}_{fg} \state{\Psi} &= 0 ~\longrightarrow ~\bar{u}^{\alpha}_{f}\frac{\partial}{\partial \bar{u}^{\alpha}_{g}} \Psi(x, \bar{u}) = 0. \label{ConOff}
\end{align}
The first constraint picks out the $F$-fold tensor product where the occupation number associated to the family $f$ is equal to $n_{f}$. This tensor product is still in general reducible, but the second constraint requires the wavefunction to be in the kernel of the $\frac{1}{2}(F^{2} - F)$ operators $\hat{L}_{fg}$. These requirements demand that the wavefunction have exactly the correct symmetry to transform under the representation whose Young Tableau has exactly $n_{f}$ rows in column $f$, so that we have the successive projections
\begin{equation}
\Yvcentermath1
\hspace{-1em}	\Psi ~ \sim \sum_{\{p_{1}, p_{2}, \ldots p_{\!F}\}}  \underbrace{\underset{ \,\Small \yng(1,1)\, }{\overset{ \,\Small\yng(1,1)\, } {\rvdots} }}_{p_{F}} \hspace{-0.5em} \otimes  \ldots \otimes \hspace{-0.5em}  \underbrace{\underset{ \,\Small\yng(1,1)\, }{\overset{ \,\Small\yng(1,1)\, } {\rvdots} }}_{p_{2}} \hspace{-0.5em}   \otimes \hspace{-0.5em}  \underbrace{\underset{ \,\Small\yng(1,1)\, }{\overset{ \,\Small\yng(1,1)\, } {\rvdots} }}_{p_{1}} \,
	\mathlarger{\overset{\hat{L}_{ff}}{\Longrightarrow }} \,
	\underbrace{\underset{ \,\Small \yng(1,1)\, }{\overset{ \,\Small\yng(1,1)\, } {\rvdots} }}_{n_{F}} \hspace{-0.5em}  \otimes  \ldots \otimes \hspace{-0.5em} \underbrace{\underset{ \,\Small\yng(1,1)\, }{\overset{ \,\Small\yng(1,1)\, } {\rvdots} }}_{n_{2}} \hspace{-0.5em}   \otimes \hspace{-0.5em}  \underbrace{\underset{ \,\Small\yng(1,1)\, }{\overset{ \,\Small\yng(1,1)\, } {\rvdots} }}_{n_{1}} \,
	 \mathlarger{\overset{\hat{L}_{fg}}{\Longrightarrow}} 	\,
	 \quad \underbrace{\overset{n_{F}...\hfill}{\underset{ \Small \yng(4,2,1) }{ \overset{ \Small \yng(4,4,4) }{ \rvdots } }} \overset{\overset{...}{\vphantom{ { \yng(1)} }}}{\underset{ \underset{...}{ \vphantom{  {\yng(1)}  } } }{...}}   \overset{\hfill ...n_{1}} {\vphantom{\underset{ \Small \yng(4,2,1) }{ \overset{ \Small \yng(4,4,4) }{ \rvdots } }} \underset{ \Small \yng(2) \hphantom{\yng(2)}  }{ \overset{ \Small \yng(4,3,2) }{ \rvdots } }}}_{F \textrm{ columns}}\, . 
	\Yvcentermath0
\end{equation}
In this way the choice of the $F$-tuple $(n_{1}, n_{2}, \ldots, n_{F})$ in the worldline action (\ref{phasecza}) ensures that the physical Hilbert space is that of a particle transforming in the irreducible representation indicated above. For further details on this construction the reader is directed to \cite{JO1, JO2}.

\section{Super-space}
We may make further progress in simplifying the worldline theory. The non-Abelian part of the field strength tensor, $-iA\wedge A$, yields a quadratic coupling to the matter fields in $\widetilde{H}$. This can be linearised with respect to $A$ by introducing additional auxiliary worldline fields $z_{\alpha}$ ($\bar{z}^{\alpha}$) with opposite statistics to the existing colour fields $c_{\alpha}$ ($\bar{c}^{\alpha}$) that will turn out to be their super-partners \cite{JO2, Paul}. To achieve this, fix $F = 1$ families of colour field and consider the term in $\psi^{\mu}\cal{G}_{\mu\nu}\psi^{\nu}$ quadratic in $A$: 
\begin{equation}
	-2i\bar{c}^{\alpha \prime}\psi^{\mu} (A_{\mu})_{\alpha \prime}{}^{\beta}(A_{\nu})_{\beta}{}^{\alpha}\psi^{\nu}c_{\alpha} \equiv  -\left(\bar{z}^{\alpha \prime} (A_{\mu})_{\alpha \prime}{}^{\alpha} \psi^{\mu} c_{\alpha} + \bar{c}^{\alpha \prime}\psi^{\mu} (A_{\mu})_{\alpha \prime}{}^{\alpha} z_{\alpha}\right)
\end{equation}
where we have defined the auxiliary variables
\begin{equation}
	z_{\alpha} \equiv i \psi \cdot A_{\alpha}{}^{\alpha \prime}c_{\alpha \prime}; \qquad \bar{z}^{\alpha} \equiv i\bar{c}^{\alpha \prime} A_{\alpha \prime}{}^{\alpha} \cdot \psi 
	\label{zzbar}
\end{equation}
to absorb the unwanted coupling in a symmetric fashion. Note these combinations also appear in the supersymmetry transformations of the colour fields (\ref{delc}). This identification can be encoded by modifying the phase space action (\ref{phasec}) to
\begin{equation}
	S[x,p,\psi, \bar{c}, c, \bar{z}, z] =\int_0^1d\tau \Big[ p\cdot \dot{x} +\frac{i}{2}\psi\cdot \dot{\psi} + i\bar{c}^{\alpha}\dot{c}_{\alpha} -e \widetilde{H}^{\prime} -i\chi \widetilde{Q}\Big]  
	\label{phasecz}
\end{equation}
where now the Hamiltonian is written in terms of the additional auxiliary fields as\footnote{\label{footQ} $\widetilde{Q}$ produces the SUSY algebra under identification of $\widetilde{H}$ with $\widetilde{H}'$ when the auxiliary variables $\bar{z}^{\alpha}$, $z_{\alpha}$ are on shell. Moreover, one can directly produce $\widetilde{H}'$ from $\{\widetilde{Q}', \widetilde{Q}'\}$ where $\widetilde{Q}^{\prime} = p\cdot \psi - \frac{i}{2}\left(\bar{z}^{\alpha} c_{\alpha} + \bar{c}^{\alpha} z_{\alpha}\right)$, using the Poisson brackets induced by (\ref{zzbar}).}
\begin{align}
	\widetilde{H}^{\prime} &= \frac{1}{2}\widetilde{\pi}^{2} + i \bar{c}^{\alpha \prime} \psi^{\mu} \partial_{\mu} (A_{\nu})_{\alpha \prime}{}^{\alpha} \psi^{\nu} c_{\alpha} + \bar{z}^{\alpha}z_{\alpha} -i\left(\bar{z}^{\alpha \prime} (A_{\mu})_{\alpha \prime}{}^{\alpha} \psi^{\mu} c_{\alpha} + \bar{c}^{\alpha \prime}\psi^{\mu} (A_{\mu})_{\alpha \prime}{}^{\alpha} z_{\alpha}\right) 
\end{align}
The new fields inherit transformations from their component fields in (\ref{zzbar}) and we find variations
\begin{align}
	\delta_{\eta} \bar{c}^{\alpha} &= i\eta \bar{z}^{\alpha} && \delta_{\eta}c_{\alpha} =i\eta z_{\alpha}	\nonumber \\
	\delta_{\eta} \bar{z}^{\alpha} &= -\frac{\eta}{e}\left(\dot{\bar{c}}^{\alpha} - i\chi \bar{z}^{\alpha}\right) && \hspace{0.15em} \delta_{\eta}z_{\alpha}  = -\frac{\eta}{e}\left(\dot{c}_{\alpha} - i \chi z_{\alpha}\right)
	\label{susycz}
\end{align} 
compatible with the Poisson brackets generated by $\widetilde{Q}$ or $\widetilde{Q}^{\prime}$ of footnote \ref{footQ}.

The extension of the local supersymmetry and reparameterisation invariance to incorporate the colour fields hints at a superspace formalism. Indeed, integrating out the canonical momenta $p^{\mu}$, (using, for instance, its equation of motion) one finds the configuration space action
\begin{align}
	\int_{0}^{1} d\tau \bigg[ \frac{1}{2}e^{-1}\dot{x}^{2} + \frac{i}{2}\psi \cdot \dot{\psi} + i\bar{c}^{\alpha} \dot{c}_{\alpha} &- \frac{i\chi}{e}\dot{x} \cdot \psi + \bar{c}^{\alpha} \mathcal{A}^{0}_{\alpha}{}^{\alpha \prime} c_{\alpha \prime}  
	 - e \bar{z}^{\alpha}z_{\alpha} + ie(\bar{z}^{\alpha} \psi \cdot A_{\alpha}{}^{\alpha \prime} c_{\alpha \prime} + \bar{c}^{\alpha} \psi \cdot A_{\alpha}{}^{\alpha \prime} z_{\alpha \prime})\bigg],
	\label{cpt}
\end{align}
where $\mathcal{A}^{0} \equiv \dot{x} \cdot A -i\psi \cdot \partial A \cdot \psi$. Note that the quadratic coupling in $\widetilde{\pi}^{2}$ has dropped out. In the general case of $F$ families of colour fields (\ref{cpt}) follows from the superspace action (extending the parameter domain $\tau \rightarrow (\tau, \theta)$ to include a Grassmann coordinate $\theta$ with super-derivative $D = \partial_{\theta} + i\theta \partial_{\tau}$) \cite{Holten, West}
\begin{equation*}
	\int\! d\tau d\theta \left[ -\frac{1}{2}\mathbf{E}^{-1} D^{2}\mathbf{X} \cdot D\mathbf{X} - \sum_{f}\bar{\mathbf{\Gamma}}_{f}^{\alpha}D\mathbf{\Gamma}_{f \alpha} + \sum_{f}i\bar{\mathbf{\Gamma}}_{f}^{\alpha} D \mathbf{X} \cdot A_{\alpha}{}^{\alpha \prime}\left(\mathbf{X}\right) \mathbf{\Gamma}_{f \alpha \prime} - \theta \sum_{f \leqslant g}a_{fg} \left( \bar{\mathbf{\Gamma}}_{f}^{\alpha}\mathbf{\Gamma}_{g \alpha} - s_{f}\delta_{fg}\right) \right],
	\label{sspace}
\end{equation*}
upon integrating over $\theta$. In the above expression we have introduced the superfields
\begin{align}
	\mathbf{X}^{\mu}\left(\tau, \theta\right) &= x^{\mu}\left(\tau\right) + \theta e^{\frac{1}{2}}\left(\tau\right)\psi^{\mu}\left(\tau\right); \qquad
	\hspace{0.2em}\mathbf{E}\left(\tau,\theta\right) = e\left(\tau\right) - 2\theta e^{\frac{1}{2}}\left(\tau\right)\chi\left(\tau\right); \nonumber \\
	\bar{\mathbf{\Gamma}}_{f}^{\alpha}\left(\tau, \theta\right) &= \bar{c}_{f}^{\alpha}\left(\tau\right) + \theta e^{\frac{1}{2}}\left(\tau\right)\bar{z}_{f}^{\alpha}\left(\tau\right); \qquad
	\mathbf{\Gamma}_{f \alpha}\left(\tau, \theta\right) = c_{f \alpha}\left(\tau\right) + \theta e^{\frac{1}{2}}\left(\tau\right)z_{f \alpha}\left(\tau\right).
	\label{sfields}
\end{align}
The transformations of the component fields follow from variations of the superfields under diffeomorphisms ($a(\tau)$ transforms as $e(\tau)$); along with (\ref{susycz}) the other supersymmetry transformations are
\begin{align}
	\delta_{\eta} x^{\mu} = i\eta \psi^{\mu}; \qquad \delta_{\eta} \psi^{\mu} = -\frac{\eta}{e}\left(\dot{x}^{\mu} - i\chi \psi^{\mu}\right); \qquad \delta_{\eta}e = 2i\eta \chi; \qquad \delta_{\eta}\chi = \dot{\eta}
\end{align}
We see that the pairs $\bar{c}^{\alpha}$ ($c_{\alpha}$) and $\bar{z}^{\alpha}$ ($z_{\alpha}$) are now completely analogous to the pair $\psi^{\mu}$ and $x^{\mu}$. Furthermore the superspace action (\ref{sfields}) demonstrates how the additional auxiliary fields linearise the interaction between matter fields and the gauge field. The $U(F)$ symmetry is also made manifest in the superspace formalism, where it appears as the obvious internal symmetry of (\ref{sspace}) under
\begin{equation}
	\bar{\mathbf{\Gamma}}^{\alpha} \rightarrow \bar{\mathbf{\Gamma}}^{\alpha}\e^{i \lambda} \qquad \mathbf{\Gamma}_{\alpha} \rightarrow \e^{-i \lambda} \mathbf{\Gamma}_{\alpha}.
\end{equation}
where $\lambda(\tau) \in u(F)$ is a generator of the symmetry. We have gauged this as in section \ref{canon}; the infinitesimal transformation of the worldline gauge fields $a_{fg}$ is then $\delta_{\lambda}a_{fg} = \dot{\lambda}_{fg} + i[a, \lambda]_{fg}$\,. 

\subsection{Path integral on $S^{1}$}
To demonstrate these techniques we compute the path integral over the colour fields appropriate for the calculation of the one-loop effective action. For this we continue to work in configuration space and Wick rotate to Euclidean metric. As in (\ref{WL}) we gauge fix the supersymmetry invariance choosing $e(\tau) = T$ and $\chi(\tau) = 0$. This leads us to the path integral for the colour fields that takes the form  
\begin{equation}
	\mathcal{Z}[A] = \frac{1}{\textrm{Vol}[{U(F)}]}\oint \mathscr{D}[\bar{c},c]\int \mathscr{D}a\, \e^{- \int_{0}^{1} d\tau \left[ \bar{c}^{\alpha}_{f} \dot{c}_{f\alpha} -i \bar{c}_{f}^{\alpha} \mathcal{A}^{a}(T^{a})_{\alpha}{}^{\alpha'} c_{f\alpha'}  + \sum\limits_{f=1}^{F}ia_{ff}\left(\bar{c}_{f}^{\alpha}c_{f\alpha}\! - \!s_{f}\right) + \sum\limits_{g < f}ia_{fg}\bar{c}_{f}^{\alpha}c_{g\alpha} \right]}
\end{equation}
over anti-periodic Grassmann functions $\bar{c}^{\alpha}_{f}$, $c_{\alpha f}$ and the worldline gauge field $a_{fg}(\tau)$ for $f \leqslant g$. We have divided by the volume of the gauge symmetry group. As we commented above, the exponential does not require path ordering since we will see that this is produced automatically by the integration over the colour fields, avoiding the need to fix the order of interactions with external gluons.

We first gauge fix by using a constant $U(F)$ rotation to set $a_{fg}(\tau) \rightarrow \hat{a}_{fg} = \textrm{diag}(\theta_{1}, \theta_{2}, \ldots, \theta_{F})$ where the $\theta_{i} \in [0, 2\pi]$ are constant angular moduli that remain to be integrated over. As shown in \cite{JO1, JO2}, the Fadeev-Popov determinant that arises from this induces a measure on the modular parameters\footnote{This measure differs from that given in \cite{JO1, JO2}, indicated by a tilde, because for convenience we have absorbed a quantum shift in the occupation numbers into it, maintaining the original definition of the $s_{f}$ in terms of un-corrected integers $n_{f}$.}
\begin{equation}
	\widetilde{\mu}\left(\{\theta_{k}\}\right) = \textrm{Det}\left(\frac{d}{d\tau} + i[a, \cdot]\right)\bigg|_{a = \hat{a}} \longrightarrow  \prod_{h < g} \left(1 - \e^{-\theta_{h}}e^{i\theta_{g}}\right).
	\label{mu}
\end{equation}
so that the functional integration over $a_{fg}(\tau)$ is replaced according to $\int \mathscr{D}a_{fg}\, \Omega[\{a_{fg}\}] \longrightarrow \textrm{Vol}[{U(F)}] \prod_{f = 1}^{F}\int_{0}^{2\pi} \frac{d\theta_{f}}{2\pi} \,\widetilde{\mu}\left(\{\theta_{k}\}\right) \Omega[\{\theta_{k}\}]$ for any functional of the worldline gauge fields. Indeed, we now find that the path integral over colour fields can be written 
\begin{align}
	 \hspace{-1.5em} \prod_{f = 1}^{F}\int_{0}^{2\pi} \frac{d\theta_{f}}{2\pi} \,\widetilde{\mu}\left(\{\theta_{k}\}\right) e^{i\theta_{f}s_{f}} \oint \mathscr{D}[\bar{c},c]\, \e^{- \int_{0}^{1} d\tau \left[ \bar{c}^{\alpha}_{f} \mathcal{D}_{f} c_{f\alpha} -i \bar{c}_{f}^{\alpha} \mathcal{A}^{a}(T^{a})_{\alpha}{}^{\alpha'} c_{f\alpha'} \right]} = \prod_{f = 1}^{F}\int_{0}^{2\pi} \frac{d\theta_{f}}{2\pi} \,\widetilde{\mu}\left(\{\theta_{k}\}\right) e^{i\theta_{f}s_{f}}  \underset{\scriptscriptstyle\rm ABC}{\textrm{Det}}{\left(i\left(\mathcal{D}_{f}- i\mathcal{A}\right)\right)}
	\label{Zfix}
\end{align}
where we introduced covariant derivatives $\mathcal{D}_{f} := \partial_{\tau} + i\theta_{f}$ (the gauge choice has diagonalised the path integral with respect to the family index $f$). Building upon \cite{Paul, MeUnif}, in \cite{JO2} the functional determinant was written in terms of traces of the super-Wilson loop in fully anti-symmetric representations:
\begin{align}
\label{Det}
	\hspace{-1em}\underset{\scriptscriptstyle\rm ABC}{\textrm{Det}}{\left(i\left(\frac{d}{d\tau} + i \theta_{f} - i\mathcal{A}\right)\right)} &= 	\det{ \left(\sqrt{e^{i\theta_{f}}W\left(1\right)} + 1/\sqrt{e^{i\theta_{f}}W\left(1\right)}\right)} \nonumber \\
	\hspace{-1em}&= \Yvcentermath1
e^{\frac{N}{2}\theta_{f}}\bigg(\tr W({ \,\boldsymbol{\cdot}  \,}) + \mathrm{tr} W( {\,\Tiny\yng(1)\, } )e^{-i \theta_{f}} + \mathrm{tr}W({\Yvcentermath1 \,\Tiny\yng(1,1)\, }) e^ {-2 i \theta_{f}} + \ldots  + \tr W(\underset{ \,\Tiny\yng(1,1)\, }{\overset{ \,\Tiny\yng(1,1)\, } {\rvdots} } )e^{-(N -1) i \theta_{f}} + \tr W({ \,\boldsymbol{\cdot}  \,})e^{-i N\theta_{f}}\bigg) 
	\Yvcentermath0
\end{align}
where 
\begin{equation}
	W(1) := \tr \mathscr{P}\e^{i \int_{0}^{1} d\tau \,\mathscr{A}^{a}(\tau)T^{a}}; \qquad \mathscr{A} = A\cdot \dot{x} - \frac{iT}{2}\psi^{\mu}G_{\mu\nu}\psi^{\nu}.
\end{equation} 
is the super-Wilson loop with the correct path ordering as previously promised and we have indicated the representation in which the trace is to be taken by its Young Tableau. Putting this into (\ref{Zfix}) we get
\begin{align*}
\mathcal{Z}[A] =	\prod_{f = 1}^{F}\int_{0}^{2\pi} \!\frac{d\theta_{f}}{2\pi}\, e^{i n_{f} \theta_{f} } &\prod_{h < g}\left(1\! -\! e^{- i \theta_{h}}e^{i \theta_{g}}\right)  \nonumber \\
	\Yvcentermath1
\times & \prod_{f = 1}^{F} \bigg(\tr W({ \,\boldsymbol{\cdot}  \,})\! + \!\mathrm{tr} W( {\,\Tiny\yng(1)\, } )e^{-i \theta_{f}} \!+\! \mathrm{tr}W({\Yvcentermath1 \,\Tiny\yng(1,1)\, }) e^ {-2 i \theta_{f}} + \ldots \! +\! \tr W(\underset{ \,\Tiny\yng(1,1)\, }{\overset{ \,\Tiny\yng(1,1)\, } {\rvdots} } )e^{-(N -1) i \theta_{f}} \! + \! \tr W({ \,\boldsymbol{\cdot}  \,})e^{-i N\theta_{f}}\bigg). 
	\Yvcentermath0
\end{align*}
It is convenient to introduce the worldline Wilson loop variables $z_{f} := \e^{\theta_{f}}$ in order to rewrite this as a series of integrals over closed contours encircling the origin in the complex plane:
\begin{equation}
	\mathcal{Z}[A] = 	\prod_{f=1}^{F}\oint \frac{dz_{f}}{2\pi i} \prod_{h < g} \left(1 - \frac{z_{g}}{z_{h}}\right) \prod_{f=1}^{F}\sum_{p_{f} = 0}^{N} \frac{\tr W_{[p_{f}]}}{z_{f}^{p_{f	} +1 - n_{f}}}
	\label{ZA}
\end{equation}
where we denote by $W_{[p_{f}]}$ the super-Wilson loop constructed from generators in the representation with $p_{f}$ fully anti-symmetric indices as enters into the bottom line of (\ref{Det}).

The crucial point is that the integration over the $U(F)$ moduli has measure (\ref{mu}) which is now responsible for projecting onto an irreducible representation. Indeed this approach is a continuous generalisation of the discrete projector used in \cite{DHoker} and the one-family projections of \cite{Bard, Col1, Col2}. These integrals impose the constraints (\ref{ConDiag}) and (\ref{ConOff}) on intermediate states in the path integral. To see how this works we consider a simple example of $F = 3$ families of colour fields with occupation numbers $(1, 2, 2)$. Then expanding the measure for three families (\ref{ZA}) provides the complex integrals
\begin{equation}
	\mathcal{Z}[A] = \oint\!\oint\!\oint \frac{dz_{1}dz_{2}dz_{3}}{(2\pi i)^{3}}\left(1 - \frac{z_{2}}{z_{1}} - \frac{z_{3}}{z_{2}} + \frac{z_{3}^{2}}{z_{1}z_{2}} + \frac{z_{2}z_{3}} {z_{1}^{2}} + \frac{z_{3}^{2}}{z_{1}^{2}}\right) \sum_{p_{1}, p_{2}, p_{3} = 0}^{N} \frac{\tr W_{[p_{1}]}\tr W_{[p_{2}]}\tr W_{[p_{3}]}}{z_{1}^{p_{1}}z_{2}^{p_{2} - 1}z_{3}^{p_{3} - 1}}.
\end{equation}
The integrals pick out the residues of the poles at $z_{i} = 0$. The final two terms in rounded brackets cannot supply a simple pole for any $p_{1}$ so immediately drop out, leaving
\begin{equation}
	\mathcal{Z}[A] = \tr W_{[2]}\tr W_{[2]}\tr W_{[1]} - \tr W_{[3]}\tr W_{[2]} - \tr W_{[3]}\tr W_{[1]}\tr W_{[1]} + \tr W_{[4]}\tr W_{[1]}
\end{equation}
where we note that $\tr W_{[0]} = 1$. By rotating onto the Cartan sub-algebra one may verify the identities (following the notation of \cite{JO2} describing representations with Young Tableaux)
\begin{align}
\Yvcentermath1
	\tr W_{[2]}\tr W_{[2]}\tr W_{[1]} &= \tr W( {\Tiny  \Yvcentermath1\yng(3,2) }) +  \tr W( {\Tiny \Yvcentermath1\yng(1,1,1,1,1) }) + 2\tr W( {\Tiny \Yvcentermath1\yng(2,1,1,1) }) + 2\tr W( {\Tiny \Yvcentermath1\yng(2,2,1) }) + \tr W( {\Tiny \Yvcentermath1\yng(3,1,1) }) \\
	 \tr W_{[3]}\tr W_{[2]} &= \tr W( {\Tiny \Yvcentermath1\yng(1,1,1,1,1) }) + \tr W( {\Tiny \Yvcentermath1\yng(2,1,1,1) }) + \tr W( {\Tiny \Yvcentermath1\yng(2,2,1) }) \\
	 \tr W_{[3]}\tr W_{[1]}\tr W_{[1]} &= \tr W( {\Tiny \Yvcentermath1 \yng(1,1,1,1,1) }) + 2\tr W( {\Tiny \Yvcentermath1\yng(2,1,1,1) }) + \tr W( {\Tiny \Yvcentermath1 \yng(2,2,1) }) + \tr W( {\Tiny \Yvcentermath1 \yng(3,1,1) }) \\
	 \tr W_{[4]}\tr W_{[1]} &= \tr W( {\Tiny\Yvcentermath1 \yng(1,1,1,1,1) }) + \tr W( {\Tiny \Yvcentermath1\yng(2,1,1,1) }) 
\Yvcentermath0
\end{align}
which are sufficient to show that (\ref{ZA}) produces the super-Wilson loop in the irreducible representation,
\begin{equation}
	\mathcal{Z}[A] = \tr W( {\Tiny  \Yvcentermath1\yng(3,2) })
\end{equation}
where the Young Tableau correctly has $n_{f}$ rows in column $f$ reading from right to left. Putting this back into the remaining path integrals over $x$ and $\psi$ we find a point particle path integral representation of the one loop effective action $(D_{\mu} := \partial_{\mu} - iA_{\mu}$) in Euclidean space-time
\begin{equation}
	\hspace{-1.5em}\Gamma[A] = 	\ln \textrm{Det}^{-\frac{1}{2}}\left[(\gamma \cdot D)^{2} + m^{2}\right] = -\frac{1}{2} \int_{0}^{\infty} \frac{dT}{T} \e^{-\frac{1}{2}m^{2}T} \oint \mathscr{D}x \mathscr{D}\psi\, \,e^{-\frac{1}{2} \int_{0}^{1} \frac{\dot{x}^{2}} {T} + \psi \cdot \dot{\psi} } \,\tr_{\mathcal{R}} \mathscr{P} \exp{\left(i\int_{0}^{1} \mathscr{A}[x(\tau), \psi(\tau)] d\tau\right)}
\end{equation}
with the trace in the representation, $\mathcal{R}$, fixed by the $(n_{1}, \ldots, n_{F})$. This provides a path integral representation of traces of path ordered exponentiated line integrals in arbitrary irreducible representations well-suited to the worldline formalism. In perturbation theory one may use worldline techniques that perserve permutation symmetry of the external gluons to avoid fixing an ordering of the gluon legs, thus combining multiple Feynman diagrams, similarly to how the fields $\psi^{\mu}$ absorb the path ordering associated to the Feynman spin factor \cite{FeynSpin} describing the spin-coupling to the background field.

\section{Conclusion and outlook}
We carried out a phase space analysis of the extended worldline action describing a Dirac spinor transforming in an arbitrary irreducible representation of $S\!U(N)$. The action is augmented by auxiliary colour fields that generate the gauge degrees of freedom. The fields maintain the worldline supersymmetry and with their accompanying super-partners we arrived at a locally supersymmetric action in superspace. Partially gauging a $U(F)$ symmetry of the colour fields allows for a projection onto an irreducible representation as we have demonstrated in a simple case. We used Grassmann colour fields but one may take them to be bosonic with minimal changes to the above construction. 

This relatively new technique has been applied to calculate scalar and spinor contributions to the gluon self energy for (anti-)symmetric representations \cite{Col1, Col2}. It has been extended to tree level to describe the scalar propagator in a non-Abelian background \cite{ColTree}, also for fully (anti-)symmetric representations. Ongoing work on the worldline approach to the spinor propagator will allow the incorporation of colour for Dirac fields. Moreover, this technique is adaptable to related problems such as generating the Lorentz index structure of the worldline approach to the one-loop Yang-Mills effective action \cite{JOPN} alongside the $U(N)$ gauge group information required in non-commutative space. 
\begin{acknowledgement}
The authors are grateful to Thierry Grandou and Ralf Hofmann for invaluable comments on the manuscript and to Christian Schubert for support, productive discussions and constructive suggestions throughout this research.
\end{acknowledgement}
 \bibliography{bibProc}
%
%


\end{document}